\documentstyle[]{article}
\begin{document}

\title{Consistent Equation of Classical Gravitation to Quantum Limit and Beyond}
\author{Shantilal G. Goradia\\
Physics Department University of Notre Dame \\
Notre Dame, IN-46556 (USA) \\
Email: Shantlal.Goradia.1@nd.edu
}
\date{}

\maketitle
\section{Introduction}
General Relativity makes a distinction between mass and space. Mass tells space how to curve and space tells mass how to move. Newtonian gravity equation makes a distinction between them by  having its numerator as mass effect and its denominator as inverse square law space effect at macroscopic approximation. At
microscopic distances it makes sense to substitute surface-to-surface distance between two nucleons for center-to-center distance between them to account for the mass space distinction, keeping in mind the smallest distance between coupled nucleons is Planck length. Any distance less than Planck makes no sense in the classical world. When we calculate the force between two nucleons of one femtometer diameter each, separated by a surface-to-surface distance of Planck length, we get the force that matches well known nuclear force i.e. $10^{40}$ times the value of the force of gravitation ``g'' calculated by assuming the Newtonian center-to-center distance of $1$ femtometer. What we get is what is described as the nuclear force in scientific literature. This leads to the question: Is the nuclear force (well recognized secondary effect of color force) high intensity gravitation?

\section{Analysis}

Consider the following equations {\bf (1)} and {\bf(2)}. The notation $d_n$ in
Equation {\bf(2)} is the diameter of the nucleons in question.
\begin{equation}
\hbox{{\bf Newtonian}}\quad F_N = Gm_1x m_2 / D^2 \label{1}
\end{equation}
\begin{equation}
\hbox{{\bf Proposed}} \quad F_P = Gm_1x m_2 / (D - d_n)^2 \label{2}
\end{equation}
The notation $D$
in the denominator in Newton's equation {\bf (1)} of the gravitational
force denotes the separating distance between the centers of mass of
the particles in question. The validity of equation {\bf (1)} has been verified
for distances as low as a few centimeters. Its validity is not verified
when the subatomic separating distance between nuclei of atoms is a few 
femtometers ({\bf fm}). Newton's equation is an approximation that
explains macroscopic observations. If Newton meant his equation to
hold true at microscopic distances, he would have explained the
binding energy. Newtonian physics implies point masses and action
between points. A point has no mass. I am asserting that, instead,
the classically deterministic Newtonian gravity originates at the
surfaces of nucleons, not at a central point within the nucleons. I
am not addressing electrons and coulomb forces in this paper. This
paper is dedicated to the investigation of the nuclear force alone at
this stage. The deviation from inverse square logic resulting from
our proposal is insignificant. The consequences of mass and space
microscopic distinction are enormous. The proposed correction is the
injection of $d_n$. Equation {\bf (2)} is good for all distances greater than
Planck length ($10^{-35}$ meters = $10^{-20}$ fm). I require the Planck
length as a lower bound so as to include the dominant first order
quantum effect in this classical model. The relative strength of the
proposed equation is the ratio obtained by dividing equation {\bf (2)} by
equation {\bf (1)}, which is
\begin{equation}
\hbox{{\bf The ratio}} \quad F_P / F_N = D^2 / (D - d_n)^2.
\end{equation}
\section{Strength of Gravity at Short Range per Eqn. {\bf (2)}}
When $D$ is very large compared
to $d_n$, $D^2$ is almost equal to $(D-d_n)^2$. The diameter of atoms is
hundreds of millions of times greater than the diameter of nucleons
located at the center of atoms. The force of gravitation calculated
by these two equations between the nucleons of two adjoining atoms is
practically the same, because the ratio $D^2 / (D-d_n)^2$ is almost equal
to one. When $D$ is small compared to $d_n$, the force of attraction
calculated by equation {\bf (2)} will be significantly greater than that
calculated by equation {\bf (1)}. The following results bring home the
concept. If we call the force of gravitation calculated by equation
{\bf (1)} ``g'', the force of gravitation calculated by equation {\bf (2)} would be
higher by the ratio $D^2 / (D - d_n)^2$. When $D$ exceeds $d_n$, by Planck
length ($10^{-20}$ fm), one obtains the ratio:\\
\\
\begin{tabular}{lcll}
$D^2/(D-d_n)^2$ & = & $(d_n + $ Planck length$)^2/(\hbox{Planck length})^2$ & (All lenghts in femtometers)\\
 & = & $(1 + 10^{-20})^2 / (10^{-20})^2$ & ($d_n = 1$ femtometer)\\
 & = & $10^{40}$. & \\
\end{tabular}
\\
\\
At surface to surface separations of 1, 2, 3, 4 and 10 femtometers, the calculated nuclear forces rapidly diminish to 4.0, 2.1, 1.77, 1.56 and 1.23 times the gravitational forces respectively and match Newtonian gravitation at 1000 femtometers as tabulated below. My calculations meet the observed boundary values. Nucleon deformation is neglected.

\begin{tabular}{|cl|c|} \hline
\multicolumn{2}{|c|} {\bf Separating Distance} & {\bf Nuclear Force / Gravity} \\ \hline \hline
One & Planck length, $10^{-20}$ fm & $10^{40}$\\ \hline
$1$ &  Femtometer & $4.0$ \\ \hline                             
$2$  & Femtometers & $2.1$ \\ \hline                            
$3$  & Femtometers & $1.77$ \\ \hline                             
$4$  & Femtometers & $1.56$\\ \hline                            
$5$  & Femtometers & $1.44$\\ \hline                            
$6$  & Femtometers & $1.36$\\ \hline                             
$7$  & Femtometers & $1.31$\\ \hline                             
$8$  & Femtometers & $1.26$\\ \hline                             
$9$  & Femtometers  & $1.23$ \\ \hline                            
$10$  & Femtometers & $1.21$ \\ \hline                            
$15$  & Femtometers & $1.15$ \\ \hline                            
$20$  & Femtometers & $1.11$ \\ \hline                            
$25$  & Femtometers & $1.09$ \\ \hline                            
$50$  & Femtometers & $1.04$ \\ \hline                            
$100$  & Femtometers & $1.02$ \\ \hline                           
$1000$  & Femtometers  & $1.00$ \\ \hline
\end{tabular}
\vspace{14in}
\section{Analogy}
Considering a hollow metal sphere containing smaller metal balls rumbling inside the sphere, a probe inside the sphere or close to the outside surface would detect non-central high intensity, indeterministic noise with intensity increasing with distance from the center. For a distant listener, the sound would be of deterministic nature originating from the center of the sphere with its intensity decreasing with distance. Indeterministic high intensity noise at a short range is deterministic low intensity sound at large distances. What this analogy brings home is that the color force is potentially the high intensity, non-central, classically indeterministic interaction. The gravity is potentially the low intensity, macroscopically central, Newtonian deterministic manifestation of the same fundamental interaction. I am taking the liberty to use the prevailing view that the nuclear force is the secondary effect of the color force to reach the following conclusion. This view does not need to be reestablished.

\section{Conjectures}
 \begin{itemize}
\item[(A)] We do not have quantum gravity: gravity is potentially not a separate fundamental interaction of Nature. 
\item[(B)] Rutherford's scattering experiments showed nuclear forces as far as $10$ femtometers \cite{2}: not that they do not exist beyond that range. At higher distances they are too weak to detect. 
\item[(C)] There is no proof of a central force detected inside the nucleons. 
\item[(D)] Despite its theoretical justification, Yukawa potential does not predict the observations.
\item[(E)] There is no feature of nuclear force that distinguishes it from gravitation. 
\item[(F)] Einstein attempted to explain nuclear force in terms of gravity \cite{1}.
\item[(G)] The Standard Model does not incorporate gravity.
\end{itemize}

\section{Conclusions}
Newtonian gravitation is potentially a deterministic manifestation of the classically indeterminate color forces addressed in QCD. The prevailing view is that the nuclear force is the secondary effect of the color force. The proposed theory connects the nuclear force with gravitation in one common equation. The combined contributions of these two clearly imply that gravitation is the Newtonian deterministic manifestation of the classically indeterminate color forces.

\section{Acknowledgements.}
I am grateful to Professor Fridolin Weber (University of Notre Dame) for his comments following the presentation of the concepts.

\end{document}